\documentclass{pasj00}
\draft

\def\lsim{\mathrel{\mathpalette\Oversim<}}
\def\gsim{\mathrel{\mathpalette\Oversim>}}
\def\Oversim#1#2{\lower0.5ex\vbox{\baselineskip0pt\lineskip0pt%
            \lineskiplimit0pt\ialign{%
          $\mathsurround0pt #1\hfil##\hfil$\crcr#2\crcr\sim\crcr}}}

\begin{document}
\SetRunningHead{}{}
%\Received{2000/12/31}%{yyyy/mm/dd}
%\Accepted{2001/01/01}%{yyyy/mm/dd}

\title{H$_2$ Line Emission 

Associated with the Formation of the First Stars}

\author{
        Hiromi MIZUSAWA \altaffilmark{1},
        Ryoichi NISHI \altaffilmark{1}
        and
        Kazuyuki OMUKAI \altaffilmark{2}}
\altaffiltext{1}{Department of Physics, Niigata University, Ikarashi, 
Niigata 950-2181}
\email{mizusawa@astro.sc.niigata-u.ac.jp}
\email{nishi@astro.sc.niigata-u.ac.jp}
\altaffiltext{2}{Division of Theoretical Astrophysics, 
National Astronomical Observatory, Mitaka, Tokyo 181-8588}
\email{omukai@th.nao.ac.jp}


\maketitle

\begin{abstract}
Molecular hydrogen line radiation emitted in formation events 
of first-generation stars are evaluated in a discussion of 
its detectability by future observational facilities.
H$_2$ luminosity evolution from the onset of prestellar collapse 
until the formation of a $\sim 100 {\rm M_{\odot}}$ protostar is followed.
Calculations are extended not only to the early phase 
of the runaway collapse but also to the later phase of accretion, 
whose observational features have not been studied before. 
Contrary to the runaway collapse phase, where the pure-rotational 
lines are always dominant, in the accretion phase 
rovibrational line emission becomes prominent.
The maximum luminosity is also attained in the accretion phase for 
strong emission lines.
The peak intensity of the strongest rovibrational line reaches 
$\sim 10^{-29}{\rm W/m^2}$, corresponding to the flux density of  
$10^{-5}{\rm \mu Jy}$, for a source at the typical redshift 
of first-generation star formation, $1+z=20$.
Although the redshifted rovibrational H$_2$ emission from 
such an epoch falls in the wavelength range of the next-generation 
infrared satellite, {\it Space Infrared Telescope for Cosmology and 
Astrophysics}, for exceeding the detection threshold $10^7$ such protostars 
are required to reach the maximum luminosity simultaneously in a 
pregalactic cloud.
It is improbable that this condition is satisfied in a realistic 
scenario of early structure formation. 
\end{abstract}

\section{Introduction}
First-generation stars can be a source of both the observed metals 
in the early universe and ultraviolet photons that cause the cosmological 
reionization.
In fact, substantial metal enrichment has been observed in 
Ly$\alpha$ clouds as early as $z\sim 5$ (Cowie et al. 1995; 
Songaila \& Cowie 1996; Songaila 2001).
Furthermore, the WMAP satellite has measured the electron scattering 
optical depth of $\tau_{e}=0.17 \pm 0.04$, which corresponds to very early 
reionization at $z\simeq 17\pm 5$ (Bennett et al. 2003). 
The growing evidence, though still indirectly, suggest that 
first-generation stars did play a significant role in the early evolution 
of galaxies.

Undisputedly the most decisive evidence for the first star formation 
is the direct detection of ongoing formation events.
First star-forming clouds are considered to be bright 
in H$_2$ line emission because the first stars are formed 
by gravitational collapse of primordial clouds, which is induced 
by H$_2$ line cooling (Saslow \& Zipoy 1967; Peebles \& Dicke 1968; 
Matsuda, Sato, \& Takeda 1969).

In this paper, we evaluate the H$_2$ line emission associated with 
first-generation star formation. 
Similar work includes that by Kamaya \& Silk (2002) and 
by Ripamonti et al. (2002). 
They, however, have followed only the evolution in the earlier phase of star 
formation, namely the runaway collapse phase, and have not extended their 
calculations well into the later phase, namely the main accretion phase. 
Here, we study the observational appearance in both evolutionary phases. 
We have found that the maximum value of H$_2$ line luminosity is indeed 
attained in the main accretion phase.

Currently, several next-generation large observational facilities, 
including {\it James Webb Space Telescope} ({\it JWST})
\footnote{http://www.jwst.nasa.gov/}, 
{\it Space Infrared Telescope for Cosmology and Astrophysics} ({\it SPICA})
\footnote{http://www.ir.isas.ac.jp/SPICA/}, 
and {\it Atacama Large Millimeter Array} ({\it ALMA}) 
\footnote{http://www.alma.nrao.edu/}, are awaited. 
One of their major scientific goals is to explore the formation and 
evolution of pregalactic objects in the high-redshift universe. 
Among them, {\it SPICA} is the most relevant to our purpose since 
it has the best sensitivity in the mid- to far-infrared ranges, 
where the strongest H$_2$ emission lines from the typical epoch for 
first star formation ($z \sim 20$) are redshifted.
In this paper, we therefore discuss the detectability of those H$_2$ line 
photons by {\it SPICA}. 

The outline of this paper is as follows: 
In section 2, the method of calculation is described. 
In section 3, the results for H$_2$ emission from a forming first star
are presented. 
Finally, in section 4, a brief summary is given and the detectability 
of first-generation stars by {\it SPICA} is discussed.

\section{The Model}
\subsection{Density and Velocity Evolution}
From the viewpoint of dynamical evolution, it is convenient to 
consider the star formation process as two distinct phases, 
i.e., before and after the birth of the central protostar.
These two phases are respectively called the runaway collapse phase 
and the mass accretion phase.
Because of their vastly different dynamical natures, we treat these two phases 
separately and use different models for describing dynamics in each phase.

\subsubsection{The Collapse Phase}
Since the typical timescale for gravitational collapse, 
the free-fall time, depends inversely on density, the denser 
region collapses faster and becomes even denser relative to the rest.
As a result, only the density in the central densest region increases, 
leaving the surrounding matter, or the envelope, almost unevolved.  
This central region has a flat density distribution with
a diameter given approximately by the central local Jeans length 
$\lambda_{\rm J}=\sqrt{c_{\rm s}^2 \pi/G\rho}$.
On the other hand, the envelope has the power-law density distribution, 
$\rho \propto r^{-2/(2-\gamma)}$ where $r$ is the distance from the center 
and  $\gamma=d{\rm ln} p/d{\rm ln}\rho$ is the effective adiabatic 
coefficient (Larson 1969).
During the runaway collapse phase, the Jeans mass 
$M_{\rm J}=\rho {\lambda_{\rm J}}^3$ decreases monotonically. 
Note, therefore, that matter that is a part of the central region but
is off-set from the center at the beginning will be left behind 
at some instance during the collapse, and will be 
incorporated thereafter into the envelope.

Considering this nature of dynamics, we use the following 
assumption for evolution in this phase:
only when the matter is inside the central region, i.e., $r<R_{\rm core}$,  
the physical quantities are assumed to evolve and the physical quantities 
outside do not evolve until the end of this phase.
The radius of the central region is given by 
$R_{\rm core}=\alpha \lambda_{\rm J}/2$, where $\alpha$ is a correction 
factor of order unity and its value will be given below.
In this way, only with the evolution of central quantities, 
we can know the density and velocity structure of the entire cloud.

The central evolution is assumed to be described by the free-fall 
relation modified by the pressure gradient:
\begin{eqnarray}
\frac{d\rho}{dt}&=&\frac{\rho}{\beta t_{\rm ff}} ,
\end{eqnarray}
where $\rho$ is the density at the center and the free-fall time, 
$t_{\rm ff}$, is 
\begin{eqnarray}
t_{\rm ff}=\sqrt{\frac{3\pi}{32G\rho}}.
\end{eqnarray}
Here, the modification due to the finite pressure gradient force is 
represented by the correction factor $\beta$.

Next we consider appropriate values for the correction factors $\alpha$, 
and $\beta$.
Omukai \& Nishi (1998; ON1998 hereafter) have studied the formation 
of the first stars 
by radiative hydrodynamics assuming the spherical symmetry and have 
found that Larson-Penston-type similarity solution for $\gamma=1.09$ 
(Yahil 1983) is a good approximation for the actual collapse dynamics.
For the Larson-Penston solutions, 
\begin{eqnarray}
\alpha&=&\sqrt{\frac{1}{\delta}} ,\\
\beta&=&\frac{1}{\sqrt{1-\delta}}, 
\end{eqnarray}
where $\delta$ is the asymptotic central value of the ratio 
of the pressure gradient force (per unit mass), $F_{\rm p}$, 
to the gravitational force, $F_{\rm g}$ for the Larson-Penston-type 
similarity solution.
The relations above can be derived from the following consideration:
The ratio $\delta$ is inversely proportional to $\alpha^2$;
\begin{equation}
\delta = F_{\rm p}/F_{\rm g} 
\simeq (\frac{P}{\rho R_{\rm core}})/(\frac{GM_{\rm core}}{R_{\rm core}^2}) 
\propto \frac{P}{G\rho^2}\frac{1}{R_{\rm core}^2}
\propto (\frac{\lambda_{\rm J}}{R_{\rm core}})^2
\propto \alpha^{-2}, 
\end{equation} 
where $P$ is the pressure and $M_{\rm core}$ is the mass in the core.
If $R_{\rm core}=\lambda_{\rm J}/2$, i.e, the diameter of the core is equal 
to the Jeans length, the pressure force exactly balances the gravity, 
i.e., $\alpha=1$.
Thus $\delta=\alpha^{-2}$.
If the pressure gradient force is neglected, the collapse timescale 
is the free-fall time $t_{\rm ff}=\sqrt{3 \pi / 32G\rho}$.
In our case, the acceleration is reduced by a factor of $1-\delta$ 
from the gravitational acceleration as a result of the pressure
force.
Then, by replacing the gravitational constant $G$ in the free-fall timescale 
to $(1-\delta)G$, the collapse timescale is $(1-\delta)^{-1/2}t_{\rm ff}$.
That is, $\beta=(1-\delta)^{-1/2}$.
For $\gamma=1.09$,  
\begin{eqnarray}
\delta=\left|\frac{\rm pressure \mbox{ }force}{\rm gravity\mbox{ } force}\right|\simeq 0.78. 
\end{eqnarray}
Then we use $\alpha=1.13$ and $\beta=2.13$.

\subsubsection{The Main Accretion Phase}
At very high density ($n>10^{20}{\rm cm^{-3}}$), because of the lack 
of an efficient cooling mechanism, the temperature rises adiabatically 
and the increased pressure gradient force eventually halts the central 
collapse.
At this stage, a hydrostatic core, or a protostar, is formed. 
The formation of the protostar marks the onset of the main accretion phase.
Although the protostar is small ($<10^{-2}{\rm M_{\odot}}$) at its birth, 
it grows by many orders of magnitude in mass due to accretion of the envelope 
matter.

After protostar formation, the pressure gradient force in the 
envelope becomes negligible in comparison with the gravitational force. 
Then we assume that the infall of the envelope matter proceeds 
according to the free-fall rate after the protostar formation.
The evolution of each mass shell is given by 
\begin{eqnarray}
\frac{dv}{dt}&=&-\frac{GM_{\rm r}}{r^2} ,\\
\frac{dr}{dt}&=&v,
\end{eqnarray} 
where $r$, $v$, and $\rho$ are the distance 
from the center, the velocity, the density of a free-falling mass shell, 
and $M_{\rm r}$ is the mass contained inside the mass shell, respectively.
In this case, the density of a mass shell increases as 
$\rho \propto r^{-3/2}$.
We calculate the evolution of each mass shell until the beginning of 
${\rm H_2}$ dissociation:
because of the increasingly short timescale and reduced H$_2$ abundance, 
time-averaged ${\rm H_2}$ emissivity during the ${\rm H_2}$ dissociation 
is only a negligible fraction of the entire emissivity.   

\subsection{Thermal and Chemical Evolution} 
The thermal evolution is followed by applying the energy equation 
to each fluid element:
\begin{eqnarray}
\frac{de}{dt}&=&-p\frac{d}{dt}\left(\frac{1}{\rho}\right)-\frac{\Lambda_{\rm cool}}{\rho} ,\\
e&=&\frac{1}{\gamma_{ad}-1}\frac{kT}{\mu m_{\rm H}} ,\\
p&=&\frac{\rho kT}{\mu m_{\rm H}} ,
\end{eqnarray}
where $e$ is the internal energy per unit mass, $\mu$ is the mean 
molecular weight and $m_{\rm H}$ is the mass of the hydrogen nucleus. 
The net cooling rate $\Lambda_{\rm cool}$ is
\begin{eqnarray}
\Lambda_{\rm cool}=\Lambda_{\rm line}+\Lambda_{\rm cont}+\Lambda_{\rm chem},
\end{eqnarray}
where $\Lambda_{\rm line}$ is the rate of cooling by ${\rm H_2}$ 
line emission, $\Lambda_{\rm cont}$ is the net rate of cooling 
by continuum emission and $\Lambda_{\rm chem}$ is the net rate of 
cooling by chemical reactions.

The line cooling rate $\Lambda_{\rm line}$ is treated as in Omukai(2000),
where the cooling rate owing to optically thick lines is 
reduced according to the escape probability formalism.
The optical depth $\tau$ of a line is evaluated locally by that 
for the shielding length
\begin{eqnarray}
l_{\rm sh}&=&\min{(\Delta S_{\rm th},\alpha\lambda_{\rm J}/2)} ; \\ 
\Delta S_{\rm th}&=&\frac{\Delta v_{\rm D}}{(dv/dr)}
=\left\{
\begin{array}{rl}
3\Delta v_{\rm D}\beta t_{\rm ff}& \mbox{(the runaway collapse phase)}\\
\sqrt{2}\Delta v_{\rm D} r^{3/2}/\sqrt{GM}& \mbox{(the main accretion phase)} ,
\end{array}
\right.
\end{eqnarray}
where $\Delta v_{\rm D}$ is the velocity dispersion.
As a source of the velocity dispersion, 
we consider only the thermal motion:, 
and then $\Delta v_{\rm D}= \sqrt{2kT/\mu m_{\rm H}}$.
We use the relation for the homogeneous collapse 
$v=r/(3\beta t_{\rm ff})$ for the runaway collapse phase and 
the free-fall velocity $v=\sqrt{2GM_{\rm r}/r}$ 
for the main accretion phase.

The net continuum cooling rate $\Lambda_{\rm cont}$ consists of 
the sum of cooling term by continuum emission and 
the heating term by continuum absorption.
Both terms are treated as in Omukai(2001; eq. B27), but, 
for the heating term, we consider the radiation field 
from the central protostar.
As mentioned below, because of the low effective temperature 
of the photosphere, $\Lambda_{\rm cont}$ contributes to cooling, 
rather than to heating, as a whole.
In our calculation, we assume that the protostellar radiation is 
the diluted black body of 6000K from the photospheric radius 
of 100$R_{\odot}$.
The photospheric radius is fixed at 100$R_{\odot}$ for simplicity.
But, in reality, it varies dramatically 
during the accretion phase, as pointed out by Omukai \& Palla (2003). 
However, the radiation field of the photosphere influences the thermal state 
in falling mass shells only moderately because of the low effective 
temperature of the photosphere, and this simple treatment does 
not affect our results significantly.  

Major chemical reactions between hydrogen compounds H, H$_2$, H$^{+}$, 
H$^{-}$, and ${\rm H_2^{+}}$ are solved.
Included are reactions 1, 2, and 7-22 in Omukai (2000) 
with the same rate coefficients.
Helium compounds are not included since they are thermally inactive 
in the temperature range in our calculation.
The chemical cooling rate $\Lambda_{\rm chem}$, which is mainly owing to 
${\rm H_2}$ formation and dissociation, 
is also treated as in Omukai (2000).

We neglect the external radiation field for simplicity.
In the accretion phase, the radiation comes from the luminous central star. 
However, in the spherical accretion, the UV radiation is not strong, 
because of the low effective temperature of the star.
Also the H$_2$ column density well exceeds the value needed for 
the self-shielding of dissociating photons.
Thus, in our model, photoionization and photodissociation are not important 
and not included.

We use an assumption similar to that in section 2.1.1: 
the physical quantities, such as temperature and chemical composition, 
of the envelope remain constant until the end of the collapse phase.
With this prescription, we can know the temperature and chemical composition 
distribution of the entire cloud as well as the density and velocity structure.
Therefore, we can infer ${\rm H_2}$-line emission from a certain position 
in the envelope using the same method described above for calculating 
the emission from the central region.
Summing the emissivity from the central region and from the envelope, 
we obtain the the ${\rm H_2}$ line luminosity of the entire star-forming core 
at each step.

\subsection{Initial Condition}
The fragmentation of metal-free gas clouds and formation of star-forming 
cores have been investigated by some authors (e.g., Uehara et al. 1996; 
Nakamura \& Umemura 2002; Abel et al. 2002; Bromm et al. 2002).
As the initial condition for our calculation, we adopt the typical values 
for the star-forming cores from Bromm et al.(2002): 
the number density  $n_{\rm H}=10^4{\rm cm}^{-3}$, temperature $T=200{\rm K}$, 
electron abundance $y({\rm e}^{-})=10^{-8}$ and H$_2$ abundance 
$y({\rm H}_2)=10^{-3}$. 
This condition corresponds to the Jeans mass of $\sim 10^3 {\rm M}_{\odot}$. 

\section{H$_2$ Luminosity from a Single Star-forming Core}
In Figure 1 the evolution of (a) the temperature and (b) the ${\rm H_2}$ 
concentration are shown against the number density.
The evolution of the central region in the collapse phase is 
shown by the solid curve.
For the accretion phase, the evolution of three representative mass 
shells is indicated.
First, we describe evolution in the runaway collapse phase:
In this phase, while the number density increases by about ten orders of 
magnitude, the temperature rises only by an order, owing to the 
radiative cooling by H$_2$ (see Fig 1 a; e.g., Palla, Salpeter and Stahler 1983; 
ON1998).
The rapid increase in the abundance of H$_2$ around 
$n_{\rm H}=10^8{\rm cm^{-3}}$ is owing to the three-body reaction 
(Palla et al. 1983). 
As a result of this, most of the hydrogen in the central region 
has been in the molecular form before the number density 
reaches $\sim 10^{12}{\rm cm^{-3}}$.
Next, let us consider the accretion phase:
In this phase, the temperature is higher and the H$_2$ 
concentration is lower than those in the runaway collapse phase 
for the same density.
The reason for this is as follows: 
During the accretion phase, the compression timescale of each mass shell 
is shorter than that in the runaway collapse phase for the same density.  
This causes a rapid temperature rise. 
As a result, the ${\rm H_2}$ dissociation starts before the formation 
of sufficient molecular hydrogen for cooling.

The four strongest H$_2$ lines in our calculation are 
rovibrational lines of 1-0Q(1), 1-0O(3), pure rotational lines 
of 0-0S(4) and 0-0S(3)\footnote
{For example, the meaning of 1-0Q(1) is as follows. 
`` 1-0 " denotes vibrational transition from $v=1$ to $v=0$, 
Q denotes that the change of rotational level, $\Delta J$, is 0.
Similarly, S and O  denote $\Delta J=-2, +2$, respectively.
The number in the bracket denotes rotational level after the transition.}, 
whose wavelengths at rest are $2.34{\rm \mu m}$, $2.69{\rm \mu m}$, 
$8.27{\rm \mu m}$, $10.03{\rm \mu m}$, respectively.
The evolution of those line luminosities is shown in 
Figure 2.
During the runaway collapse phase, pure-rotational lines are stronger than 
rovibrational lines. 
In contrast to this, in the accretion phase, the rovibrational lines become 
stronger than pure-rotational lines because of the high temperature and 
density in this phase.
Note that the luminosities of both rovibrational and pure-rotational 
lines reach the maximum in the accretion phase.
The peak values of the line luminosities $L_{\rm peak}$ 
and the mass of the protostar at this epoch $M_{\rm peak}$ 
are shown in Table I.

\begin{center}
\begin{tabular}{c|c|c|c}
\multicolumn{4}{c}{\bf Table I. maximum line luminosities}\\ \hline\hline
line  &$\lambda$(${\rm \mu m}$)  &$L_{\rm peak}$($10^{35}$erg/s)  
&$M_{\rm peak}$(${\rm M}_\odot$) \\ \hline
1-0Q(1)  &2.34  &1.73  &12.9  \\
1-0O(3)  &2.69  &1.71  &11.2  \\
0-0S(4)  &8.27  &0.78  &7.4  \\
0-0S(3)  &10.03 &1.05  &10.5  \\
\hline
\end{tabular}
\end{center}

Here we discuss why the line luminosities are stronger in the accretion 
phase than in the collapse phase.
The possible sources of energy radiated by the ${\rm H_2}$ emission 
include the gravitational energy liberated by the infall and the 
chemical binding energy liberated by ${\rm H_2}$ formation.
Before the peak of luminosity, the main source is the latter.
The majority of the former turns into kinetic energy without 
being converted to radiation energy.
On the other hand, most of the chemical binding energy liberated by 
${\rm H_2}$ formation is converted into thermal energy on the spot, 
since the number density is high enough to assure the local thermodynamic 
equilibrium ($n \gg 10^{4}{\rm cm^{-3}}$) in our calculation.
The evolution of the liberation rate of chemical energy and the H$_2$ 
radiation emissivity for the mass shell that contains about 
$13{\rm M_{\odot}}$ inside are shown in Figure 3. 
Hereafter we write this mass shell as ``mass shell ($13{\rm M_{\odot}}$)''.
As is seen in the figure, the majority of the radiated energy is indeed
coming from the liberated chemical binding energy.
We compare here the liberation rate of chemical energy in the 
collapse phase with that in the accretion phase.
In Figure 4, we show the chemical energy liberated per 
unit time at four selected instances, the final stage of the collapse 
phase and three steps of the accretion phase, against the radial distance.
The liberation rate of chemical energy is higher in the accretion phase
than in the collapse phase.
The reason for this is as follows:
The radius of the H$_2$ forming region, which is defined as the radius 
where the ${\rm H_2}$ formation rate is maximum, decreases from
$\sim 3\times 10^{15}({\rm cm})$ in the collapse phase to  
$\sim 6\times10^{14}({\rm cm})$ when the mass of the central protostar 
is $\sim 5{\rm M_{\odot}}$ in the accretion phase. 
At the same time the inflow velocity and thus the mass entering 
into the ${\rm H_2}$ formation region per unit time becomes larger 
in the accretion phase, i.e. the total number of H$_2$ forming per 
unit time increases.
Since the liberated chemical binding energy is converted into 
radiation efficiently, the luminosity in the accretion phase 
exceeds that in the collapse phase. 

The luminosities of rovibrational lines reach their maximum at a 
protostellar mass of about $13{\rm M}_\odot$ (Figure 2). 
Here, we discuss why this luminosity peak appears.
In Figure 5, we show the evolution of the mass shell ($13{\rm M}_\odot$) 
inside, against the radial distance. 
For comparison, we also plot some of the other mass shells.
When the mass of the protostar is small, 
despite the large emissivity per unit mass, the mass of the region 
appropriate for vibrational lines, where the temperature is between 1000K and 
2000K, is small. 
On the other hand, at a later phase, owing to the fast contraction 
due to the strong gravity of the central protostar, the temperature 
rapidly reaches the H$_2$ dissociation value ($\simeq 2000$K) before 
sufficient H$_2$ formation.
The radiative emissivity is smaller, although the mass of the emission region 
is larger. 
Consequently, the luminosity peak appears at some epoch in between.

In the outer envelope where density is low 
($\sim 10^{6}-10^{12}{\rm cm^{-3}}$), the temperature of infalling 
mass shells in the collapse phase is higher by a factor of $\lesssim 2$
than the value of the Larson-Penston-type similarity solution for 
$\gamma =1.09$.
Therefore, the mass accretion rate, which depends on the temperature 
as ${T^{\frac{3}{2}}}$, $dM/dt \sim c_{\rm s}^3/G \propto T^{3/2}$
(e.g., Stahler et al. 1986), becomes larger than 
that derived from the similarity solution by a factor of $\lesssim 4$ 
after the protostellar mass exceeds $\sim 3{\rm M_{\odot}}$.
The mass accretion rate so calculated decreases slowly from the initial 
value of $\sim 0.05{\rm M_{\odot}/yr}$ 
to $\sim 0.02{\rm M_{\odot}/yr}$ at the protostar mass 
of $100 {\rm M_{\odot}}$.

\section{Discussion and Conclusion}
We have studied the evolution of H$_2$ line emission from a forming
first star, starting from a prestellar dense core up to the formation of 
a $100{\rm M_{\odot}}$ protostar.
After the formation of the central protostar, the H$_2$ line luminosity 
increases and luminosities of strong lines reach their maximum 
when the central protostellar mass is $\sim 13{\rm M_{\odot}}$. 
The four strongest ${\rm H_2}$ lines are the rovibrational 
lines of  1-0Q(1) ($2.34{\rm \mu m}$), 1-0O(3) ($2.69{\rm \mu m}$), 
the pure-rotational lines, 0-0S(4) ($8.27{\rm \mu m}$), 
and 0-0S(3) ($10.03{\rm \mu m}$). 
During the accretion phase, vibrational lines of 1-0Q(1) and 
1-0O(3) are the strongest, while in the collapse phase the strongest 
lines are the pure-rotational lines. 

Strong H$_2$ line emission is expected also from diffuse 
regions in low-metallicity ($Z < 0.01Z_{\odot}$) forming galaxies 
(e.g., Shchekinov \& \'{E}nt\'{e}l' 1985; Shchekinov 1991; 
Omukai \& Kitayama 2003). 
However, the dominant lines from those sources are pure-rotational 
lines because of the lower density ($n \sim 10^{0-3} {\rm cm^{-3}}$).
On the other hand, in the late stage of the first star formation, 
${\rm H_2}$ rovibrational lines are dominant, because of 
both high density ($n \sim 10^{10-12} {\rm cm^{-3}}$) and
temperature ($T> 1000$K) in the H$_2$ emitting region.
Therefore, detection of ${\rm H_2}$ rovibrational lines can be 
a useful method to identify first-generation star forming clouds.

Several large observational projects using next-generation facilities 
are now in progress, including {\it JWST}, {\it SPICA} and {\it ALMA}. 
The wavelength ranges of these three facilities are presented in Table II.
\begin{center}
\begin{tabular}{c|c|c}
\multicolumn{3}{c}{\bf Table II. The wavelength ranges 
of future telescopes}\\ \hline\hline
{\it JWST}      & {\it SPICA}    & {\it ALMA} \\
\hline
$0.6-28{\rm \mu m}$  &$5-200{\rm \mu m}$  &$300{\rm \mu m}-3{\rm mm}$\\
\hline
\end{tabular}
\end{center}
For typical formation epochs of the first stars $z\sim 20$ 
(e.g., Abel et al. 2002), 
the emitted H$_2$ rovibrational line photons are redshifted to 
the far-infrared wavelength. 
As is seen in Table II, {\it SPICA} has the deepest detection limit 
in this range and thus is the most suitable to our purpose. 
{\it SPICA} is a next-generation infrared satellite, which is planned to 
have a cooled (4.5K) and large (3.5m) telescope and to be highly 
sensitive at the mid- and far-infrared wavelengths.
The line detection limit of {\it SPICA} is about $10^{-20}{\rm W/m^2}$ 
in the far-infrared wavelength $\lambda\sim 40{\rm \mu m}$ with
a spectroscopic resolution $R=3000$ and observation time of one hour. 
Since a signal to noise ratio is proportional to $\sqrt{t}$, 
the sensitivity improves, of course, with longer integration time.
Nonetheless, in the mid- and far-infrared wavelengths, 
the high background of the zodiacal light limits the sensitivity 
of {\it SPICA} to $10^{-(21-22)}{\rm W/m^2}$
 (private communication H. Matsuhara).

The observed wavelengths of the four strongest lines redshifted 
from the emission at $1+z=20$, the typical epoch of first star formation, 
are given in Table III.
\begin{center}
\begin{tabular}{l|c|c|c|c}
\multicolumn{5}{c}{\bf Table III. The Observed Wavelengths of 
Strong H$_2$ Lines}\\ \hline\hline
line & 1-0Q(1) & 1-0O(3) & 0-0S(4)  & 0-0S(3) \\
\hline
rest frame &$2.34{\rm \mu m}$  &$2.69{\rm \mu m}$  &$8.27{\rm \mu m}$  
&$10.03{\rm \mu m}$ \\
\hline
redshifted ($1+z=20$) &$46.8{\rm \mu m}$  &$53.8{\rm \mu m}$  
&$165.4{\rm \mu m}$  &$200.6{\rm \mu m}$\\
\hline
\end{tabular}.
\end{center}
The wavelengths of the redshifted pure-rotational lines shown above are 
outside the range of any future telescope given in Table III.
On the other hand, the wavelengths of the rovibrational lines, 
which are intrinsically stronger than pure-rotational lines, 
fall in the range where the sensitivity of {\it SPICA} is excellent.

By using the value of peak luminosity $L_{\rm peak}$ in Table I,  
the peak intensity of the strongest rovibrational lines 
from $1+z=20$ is
\begin{eqnarray}
F_{\rm peak}=\frac{L_{\rm peak}}{4\pi D_{1+z=20}^2}
=\frac{1.73\times10^{35}{\rm erg}/{\rm s}}{4\pi(2.13\times 10^{11}{\rm pc})^2}
\simeq 0.32 \times 10^{-28}{\rm W}/{\rm m}^2,
\end{eqnarray}
where $D_{1+z=20}=2.13\times 10^{11}{\rm pc}$ is the 
luminosity distance to $1+z=20$ given by the formula 
\begin{eqnarray}
D_{1+z}=\frac{c (1+z)}{H_0}\int_{1}^{1+z}
\frac{dy}{[\Omega_{\Lambda}+\Omega_{\rm M}y^{3}]^{1/2}},
\end{eqnarray}
where $H_0$ is the Hubble parameter, with the values 
$H_0=70{\rm km}{\rm s}^{-1}{\rm Mpc}^{-1}$, $\Omega_{\Lambda}=0.7$, 
$\Omega_{\rm M}=0.3$.
This peak intensity corresponds to the peak flux density 
of $\sim 10^{-5}{\rm \mu Jy}$ using the line width $\nu_0 v/c 
\sim 10^{10}{\rm Hz}$, where $\nu_0$ is the line-center frequency 
for the observer and $v$ is the velocity of the mass shell. 
Note that in considering a cluster of star-forming cores, 
a larger value of the line width should be taken because 
relative velocity between the cores is usually larger than 
the thermal velocity in a core. 
In this case, the peak flux density per unit core will be 
lower than the value given above.

Taking into account the limiting sensitivity 
$10^{-(21-22)}{\rm W/m^2}$ and the maximum rovibrational line flux 
$\sim 10^{-29} {\rm W/m^2}$, we conclude that 
{\it SPICA} is able to observe a cluster of forming first-generation stars 
if it is contains more than $10^{7}$ sources with their 
luminosity near the peak value. 

Next, we consider whether or not this situation is possible in a 
realistic cosmological context.  
The epoch of the peak luminosity lasts only for 
$\Delta t_{\rm peak} \sim 10^{3} {\rm yrs}$ (see Fig 2), 
which is by far shorter than the typical free-fall time in the 
pregalactic clouds (say, $t_{\rm ff} \gsim 10^{7}$yrs).
Even in the extreme case of starburst, star 
formation is a prolonged event that continues at least 
a free-fall timescale of the pregalactic cloud.
Thus, at a given time, at most only 
$\Delta t_{\rm peak} /t_{\rm ff} \lsim 10^{-4}$ 
of the total gas mass of the pregalactic cloud is a total mass of 
star forming cores approximately at peak luminosity.
To reach the luminosity peak, the first-star forming cores must be 
as massive as $13{\rm M_{\odot}}$ (\S 3).
The mass scale of actual cores is probably more massive, typically 
$10^{2-3}{\rm M_{\odot}}$, according to numerical simulations of 
the fragmentation of primordial clouds (e.g., Bromm et al. 2002).
Thus the total mass of 10$^{7}$ cores is $\sim 10^{9}{\rm M_{\odot}}$.
Since only $10^{-4}$ of all the cores are at peak luminosity, 
in order for 10$^{7}$ cores to reach peak luminosity at the same time, 
the total cloud mass must be as high as $10^{13}{\rm M_{\odot}}$.
Note also that in this case the star formation rate is required to 
be an improbably high value of $10^6{\rm M_{\odot}/yr}$.
It is true that for pure-rotational lines, i.e., 0-0S(3), 
the epoch in which the luminosity is higher than $10^{34}$erg/s 
continues for as long as $10^5$yrs.
Although this long emission period reduces the required 
star formation rate, it is still improbably high 
($10^5{\rm M_{\odot}/yr}$).
On the other hand, in order to emit radiation mainly in H$_2$ lines during 
the prestellar collapse, especially for the very high density region, 
the metallicity in the pregalactic clouds 
must be lower than $\sim 10^{-4}Z_{\odot}$ (Omukai 2000).
However, formation of Galactic scale objects with such low metallicity is 
probably excluded in the context of standard cosmology 
(Scannapieco, Schneider, \& Ferrara 2003).
In conclusion, it is improbable that detection of H$_2$ line 
emission from forming first stars can be accomplished with {\it SPICA}.

In this paper, the accretion flow has been assumed to be spherical.
In reality, however, an accretion disk should form owing to an angular 
momentum barrier.
In this case, the total energy emitted in ${\rm H}_2$ lines might 
increase, since the released gravitational energy is converted 
more efficiently into radiation in the accretion disk.
At the same time, H$_2$ is more susceptible to photodissociation 
owing to direct irradiation from the star.
To estimate the H$_2$ luminosity more precisely, it is necessary to study
effects of the disk in the accretion phase.
We leave this problem for future work.
Recently, Bromm \& Loeb(2004) performed a three-dimensional 
numerical simulation and derived the mass accretion rate onto 
the protostar.
Their accretion rate is similar to our value 
until $\sim 10^3$ yr after the protostar formation.
After this time, their accretion rate drops more rapidly 
than ours. 
This difference is probably caused by more realistic evolution 
that they calculated, 
e.g. quasi-static contraction in the early evolution and/or 
the complex accretion flow owing to 3D effects etc.
In our calculation the maximum values of the H2 luminosities 
are reached at $\sim 300$ yr after the protostar formation.
At this epoch, our accretion rate does not differ much 
from their value.
Therefore, the peak value of H2 luminosity is expected to be almost 
the same even if their accretion rate is adopted.
After $\sim 10^3$ yr of protostar formation, we are probably 
overestimating the accretion rate and thus the H2 luminosity.
Nevertheless, the H2 luminosity at this time is already weak and 
does not important for our discussion on observability by future 
facilities.

Contrary to our claim, Kamaya \& Silk (2002) and Ripamonti et al.(2002)
have concluded that future observational facilities, e.g., {\it JWST}, 
{\it ALMA}, {\it ASTRO-F}, are able to detect H$_2$ emission from the first 
star-formation sites. 
Here we stress that our estimated value of the peak luminosities 
of ${\rm H_2}$ lines is larger than their value, which is attained for 
the earlier phase.
Moreover, in the mid- to far-infrared wavelengths,
the sensitivity of {\it SPICA} that we use here to estimate 
the detectability exceeds that of the other facilities 
that these authors considered. 
The reason for the apparent contradiction is as follows: 
in converting the emitted H$_2$ luminosity to the observed flux, 
they used the luminosity distance underestimated by a factor of $1+z$, 
where $z$ is the redshift of the source.
This error propagates to the final value of the observed flux, which 
is overestimated by $(1+z)^2$.
Note that, anyway, our value of the intrinsic luminosity in the runaway
collapse phase is consistent with Kamaya \& Silk (2002) and 
Ripamonti et al.(2002).

Finally, we touch on another future telescope, 
{\it the Single Aperture Far-Infrared Observatory} ({\it SAFIR})
\footnote{ http://safir.jpl.nasa.gov/whatIs/index.asp}.
Thanks to its larger mirror size (8 m), the line detection limit of
{\it SAFIR} is better than that of {\it SPICA} and the background limit 
of $\sim 10^{-22}{\rm W/m^2}$ at $\lambda \sim 40{\rm \mu m}$ 
can be reached in one hour with spectroscopy of $R=1000$.
Recall that the detection limit we used in this paper is also the 
background limit $\sim 10^{-22}{\rm W/m^2}$, which is set by the 
zodiacal light. 
Therefore, even with {\it SAFIR} we conclude that detecting 
${\rm H}_2$ line emission from forming first stars is difficult.

The authors would like to thank Tetsu Kitayama, Hajime Susa and 
Masayuki Umemura for valuable comments, and Ken-ichi Oohara and
Kazuya Watanabe for continuous encouragement.
This work was supported in part by the Ministry of Education, Culture,
Sports, Science, and Technology of Japan, Grants-in-Aid for Scientific 
Research No. 14540227 (RN) and Research Fellowship of the Japan Society 
for the Promotion of Science for Young Scientists, grant 6819 (KO).

\clearpage
\newpage

\begin{figure}
  \begin{center}
    \FigureFile(80mm,80mm){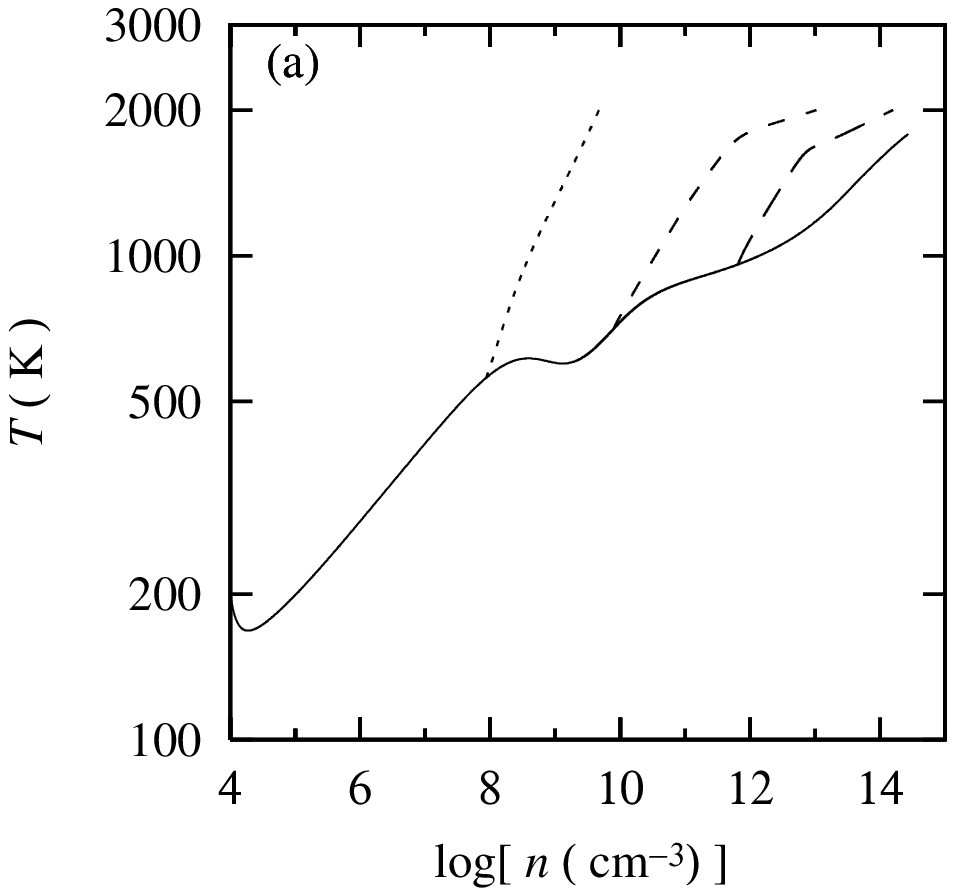}
    \FigureFile(80mm,80mm){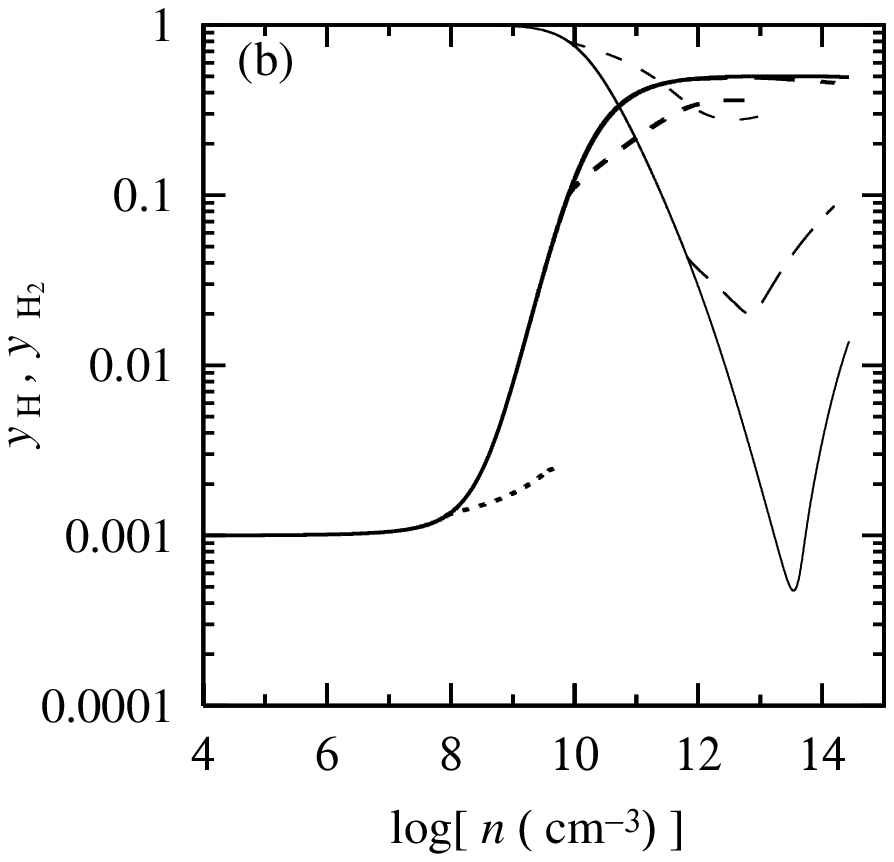}
  \end{center}
\caption{(a) The temperature evolution and 
(b) the evolution of the H$_2$ (bold line) and H (thin line) concentration 
against the number density. 
For the runaway collapse phase, the evolutionary trajectory of the central 
part is drawn with a solid line. 
For the accretion phase, the evolutionary trajectories of the mass 
shells of $M/ {\rm M}_{\odot} = $ 1(long-dashed lines), 
13(short-dashed lines) and 100(dotted lines) are shown.}
\label{fig:TH2}
\end{figure}

\newpage

\begin{figure}
  \begin{center}
    \FigureFile(120mm,80mm){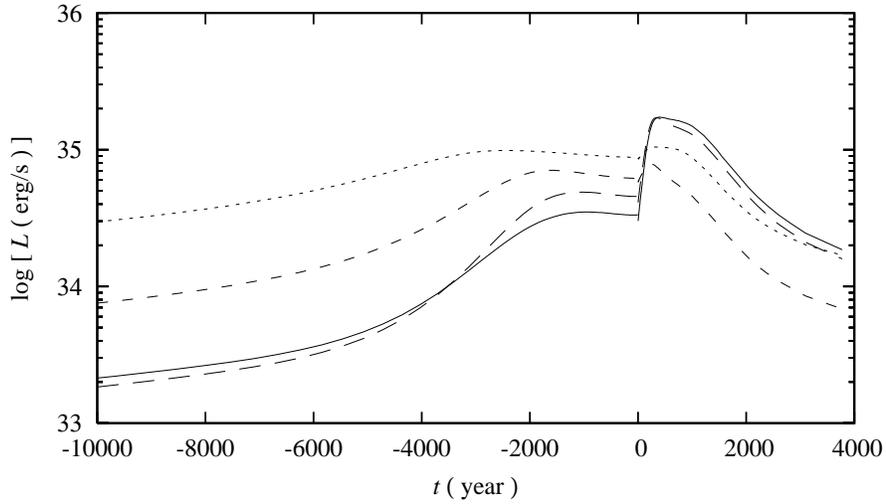}
  \end{center}
\caption{The luminosity evolution of the four strongest 
${\rm H_2}$ lines, the rovibrational lines of 1-0Q(1) (solid lines), and
1-0O(3) (long-dashed lines), and the pure rotational lines of 0-0S(4) 
(short-dashed lines), and 0-0S(3) (dotted lines). 
The horizontal axis is the time since the onset of the collapse.
The evolution up to the formation of a $10^2 {\rm M_{\odot}}$ protostar is 
shown. 
The formation of the central protostar corresponds to $t=0$.
Note that all lines attain their maximum luminosity at $t>0$, 
i.e., in the main accretion phase. 
Also the luminosities of rovibrational lines become stronger than  
that of pure-rotational ones in this phase. }
\label{fig:L(time)2}
\end{figure}

\newpage

\begin{figure}
  \begin{center}
    \FigureFile(80mm,60mm){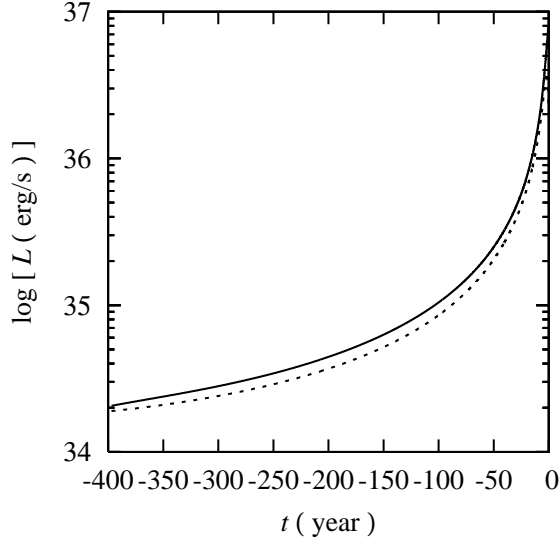}
  \end{center}
\caption{The evolution of the liberation rate of chemical binding energy 
(dotted line) and H$_2$ line emissivity (solid line) for 
the mass shell (13${\rm M}_{\odot}$). 
The time $t=0$ is set when the mass shell (13${\rm M}_{\odot}$) is 
supposed to reach the center ($r=0$). 
Because of the finite radius of the star, the mass shell reaches the 
stellar surface before $t=0$.}
\label{fig:energy}
\end{figure}

\newpage

\begin{figure}
  \begin{center}
    \FigureFile(120mm,80mm){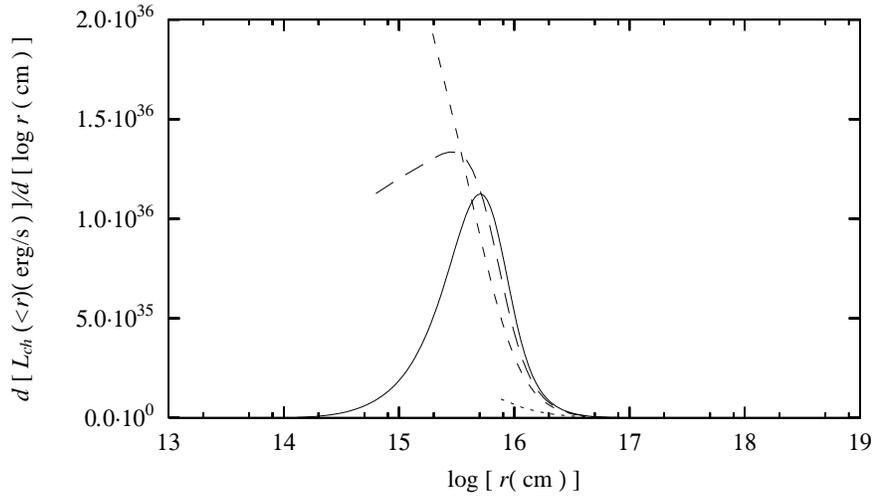}
  \end{center}
\caption{Liberation rate of chemical binding energy, 
at the final stage of the collapse phase (solid line), 
when the mass shells of $M_{\rm r}/ {\rm M}_{\odot} = $ 5 (long-dashed line), 
13 (short-dashed line) and 50 (dotted line) fall on the center, respectively.
The horizontal axis is the distance from the center. 
The area surrounded by each line and the horizontal axis 
indicates the chemical energy liberated per unit time.}
\label{fig:energy}
\end{figure}

\newpage

\begin{figure}
  \begin{center}
    \FigureFile(80mm,40mm){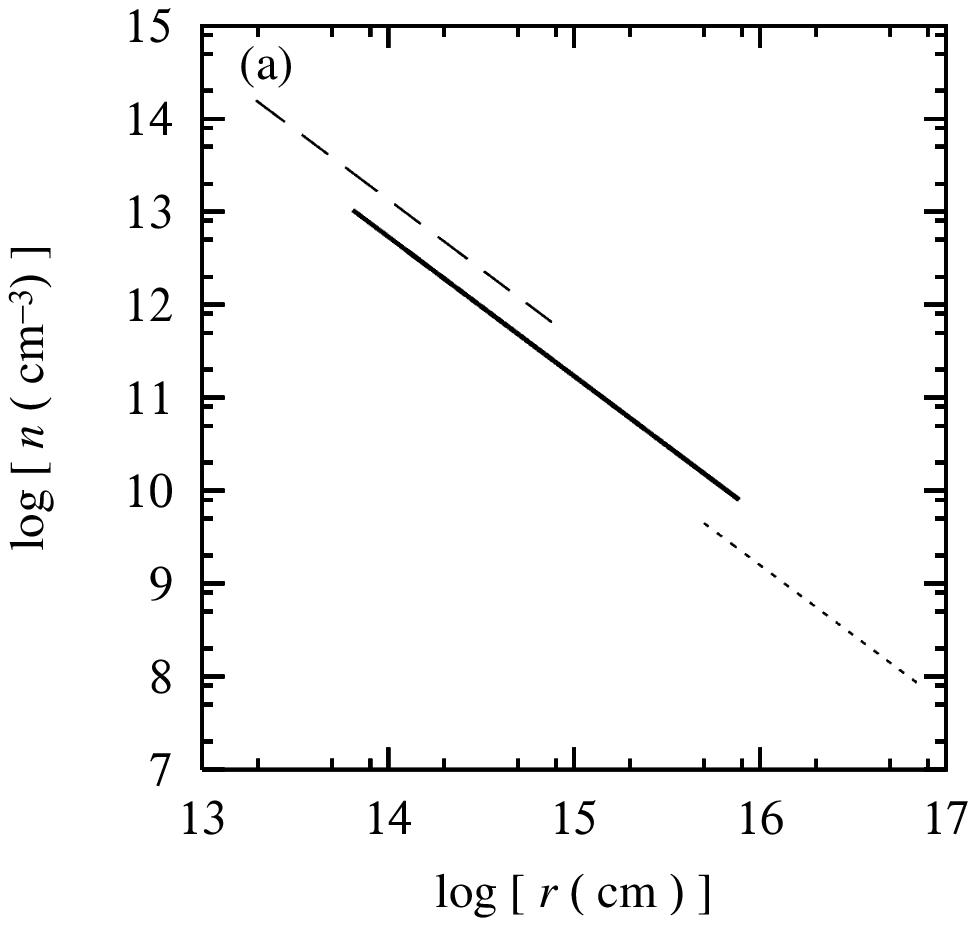}
    \FigureFile(80mm,40mm){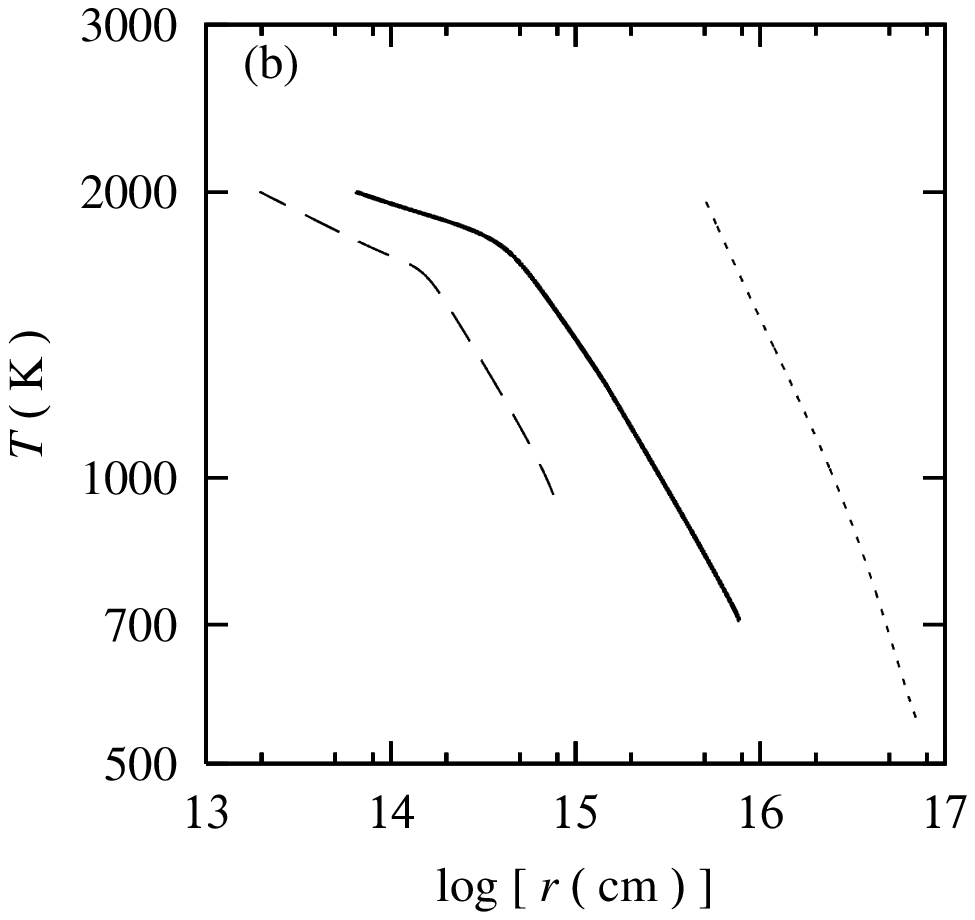}
    \FigureFile(80mm,40mm){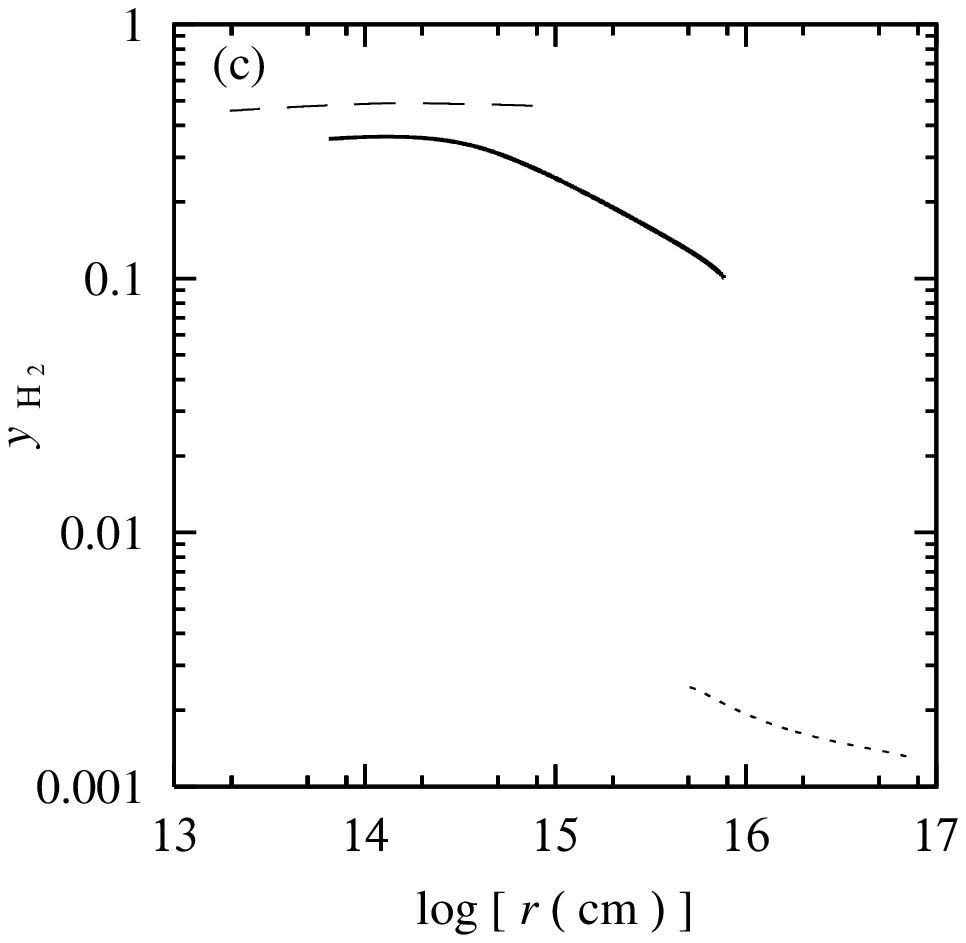}
  \end{center}
\caption{The evoluton of (a) number density, (b) temperature, 
(c) ${\rm H_2}$ concentration of selected mass shells against 
the radial distance from the central protostar in the main accretion
phase.
The values for the mass shell ($13{\rm M}_{\odot}$) are shown by the 
thick solid line.
Recall that the radii of the mass shells decrease with time  
since the envelope matter is falling onto the protostar.
For comparison, we also plot the values of other two mass shells 
that contain $1{\rm M}_{\odot}$ (dashed line) and $100{\rm M}_{\odot}$ 
(dotted line) inside. }
\label{fig:nTH2}
\end{figure}

\clearpage
\newpage
%%%%%%%%%%%%%%%%%%%%%%%%%%%%%%%%%%%%%%%

\end{document}